\documentclass[lettersize,journal]{IEEEtran} 
\usepackage{cite}
\usepackage{mathrsfs}
\usepackage{amsthm}
\usepackage[cmex10]{amsmath}
\usepackage{mathtools}
 \usepackage{siunitx}
  \usepackage{bbding}
\usepackage{array}
\usepackage{etoolbox} \makeatletter \patchcmd{\@makecaption} {\scshape} {} {} {} \makeatother
\usepackage[longtable]{multirow}
\usepackage{longtable}
\usepackage{booktabs}
\usepackage{eqparbox}
\usepackage{bm}

\usepackage[table,dvipsnames]{xcolor}
\usepackage{multicol,booktabs,tabularx}

\usepackage{cases}
\usepackage{algorithm}
\usepackage{paralist}
\usepackage{algpseudocode}

\usepackage{graphicx}
\usepackage{subfigure}
\usepackage{epstopdf}
\usepackage{epsfig}
\usepackage{amssymb}
\usepackage{array}
\usepackage{multirow}

\def\BibTeX{{\rm B\kern-.05em{\sc i\kern-.025em b}\kern-.08em
    T\kern-.1667em\lower.7ex\hbox{E}\kern-.125emX}}
\title{Enabling 6G Through Multi-Domain Channel Extrapolation: Opportunities and Challenges of Generative Artificial Intelligence}
\author{Yuan Gao, Zichen Lu, Yifan Wu, Yanliang Jin, Shunqing Zhang, Xiaoli Chu, \\Shugong Xu, \textit{Fellow, IEEE}, Cheng-Xiang Wang, \textit{Fellow, IEEE}
\thanks{This work was supported in part by Shanghai Natural Science Foundation under Grant 22ZR1422200, and in part supported by the 6G Science and Technology Innovation and Future Industry Cultivation Special Project of Shanghai Municipal Science and Technology Commission under Grant 24DP1501001. (Shugong Xu is the corresponding author)}
\thanks{Yuan Gao, Zichen Lu, Yifan Wu, Yanliang Jin and Shunqing Zhang are with the School of Communication and Information Engineering, Shanghai University, China, email: gaoyuansie@shu.edu.cn, luzichen@shu.edu.cn, 22721189@shu.edu.cn, jinyanliang@staff.shu.edu.cn and shunqing@shu.edu.cn.}
\thanks{Xiaoli Chu is with the Department of Electronic and Electrical Engineering, the University of Sheffield, UK, email: x.chu@sheffield.ac.uk.}
\thanks{Shugong Xu is with Xi’an Jiaotong-Liverpool University, Suzhou, China, email: shugong.xu@xjtlu.edu.cn.}
\thanks{Cheng-Xiang Wang is with National Mobile Communications Research Laboratory, School of Information Science and Engineering, Southeast University, Nanjing, China, e-mail: chxwang@seu.edu.cn}
}

\begin{document}
\maketitle
\begin{abstract}
Channel extrapolation has attracted wide attention due to its potential to acquire channel state information (CSI) with high accuracy and minimal overhead. This is becoming increasingly crucial as the sixth-generation (6G) mobile networks aim to support complex scenarios, for example, high-mobility communications utilizing ultra-massive multiple-input multiple-output (MIMO) technologies and broad spectrum bands, necessitating multi-domain channel extrapolation. Current research predominantly addresses channel extrapolation within a single domain, lacking a comprehensive approach to multi-domain channel extrapolation. To bridge the gap, we propose the concept of multi-domain channel extrapolation, detailing the essential performance requirements for 6G networks. These include precise channel extrapolation, adaptability to varying scenarios, and manageable computational complexity during both training and inference stages. In light of these requirements, we elaborate the potential and challenges of incorporating generative artificial intelligence (GAI)-based models for effective multi-domain channel extrapolation. Given the ability of the Transformer to capture long-range dependencies and hidden patterns, we propose a novel Transformer encoder-like model by eliminating the positional encoding module and replacing the original multi-head attention with a multilayer perceptron (MLP) for multi-domain channel extrapolation. Simulation results indicate that this model surpasses existing baseline models in terms of extrapolation accuracy and inference speed. Ablation studies further demonstrate the effectiveness of the module design of the proposed design. Finally, we pose several open questions for the development of practical GAI-based multi-domain channel extrapolation models, including the issues of explainability, generalization, and dataset collection.

\end{abstract}
\begin{IEEEkeywords}
6G, Multi-domain channel extrapolation, generative artificial intelligence, Transformer.
\end{IEEEkeywords}
\section{Introduction}

Accurate channel state information (CSI) is crucial for the optimal performance of mobile networks and has been the subject of extensive research \cite{wang2023road}. Conventional methods for acquiring CSI, such as channel estimation, require significant radio resources. To address the overhead associated with CSI acquisition, channel extrapolation techniques have been developed to infer the CSI of interest from known CSI \cite{zhang2023ai}.
\begin{itemize}
\item In ultra-high-speed communication scenarios, including vehicle-to-everything (V2X) systems, high-speed trains (HST), and unmanned aerial vehicles (UAVs), channel aging poses a significant challenge due to the rapid shifts in the wireless environment \cite{jiang2022accurate}. As a result, CSI acquisition must be performed more frequently, contributing to substantial overhead. Time-domain channel extrapolation leverages historical CSI to predict future CSI based on temporal correlations, thereby minimizing the frequency of CSI estimations \cite{jiang2022accurate}.

\item To improve mobile network performance in areas such as throughput, sensing, and positioning, it is essential to utilize multiple spectrum bands, ranging from sub-\SI{6}{\giga\hertz} to terahertz (\SI{}{\tera\hertz}). Frequency-domain channel extrapolation enables the derivation of CSI for a target band based on known CSI from another band, effectively reducing overhead in systems utilizing multiple bands or frequency-division duplex (FDD) configurations \cite{zhang2021deep}.

\item Ultra-massive multiple-input multiple-output (MIMO) technology is expected to play a pivotal role in the evolution of mobile networks, providing enhanced spectral efficiency, superior interference mitigation, and increased network capacity. To further alleviate the overhead related to CSI acquisition, space-domain channel extrapolation is proposed to extrapolate CSI for one antenna set by utilizing known CSI from a different antenna set\cite{zhang2022deep}.
\end{itemize}

\begin{figure*}[htbp]\centering    
\includegraphics[width=1.8\columnwidth]{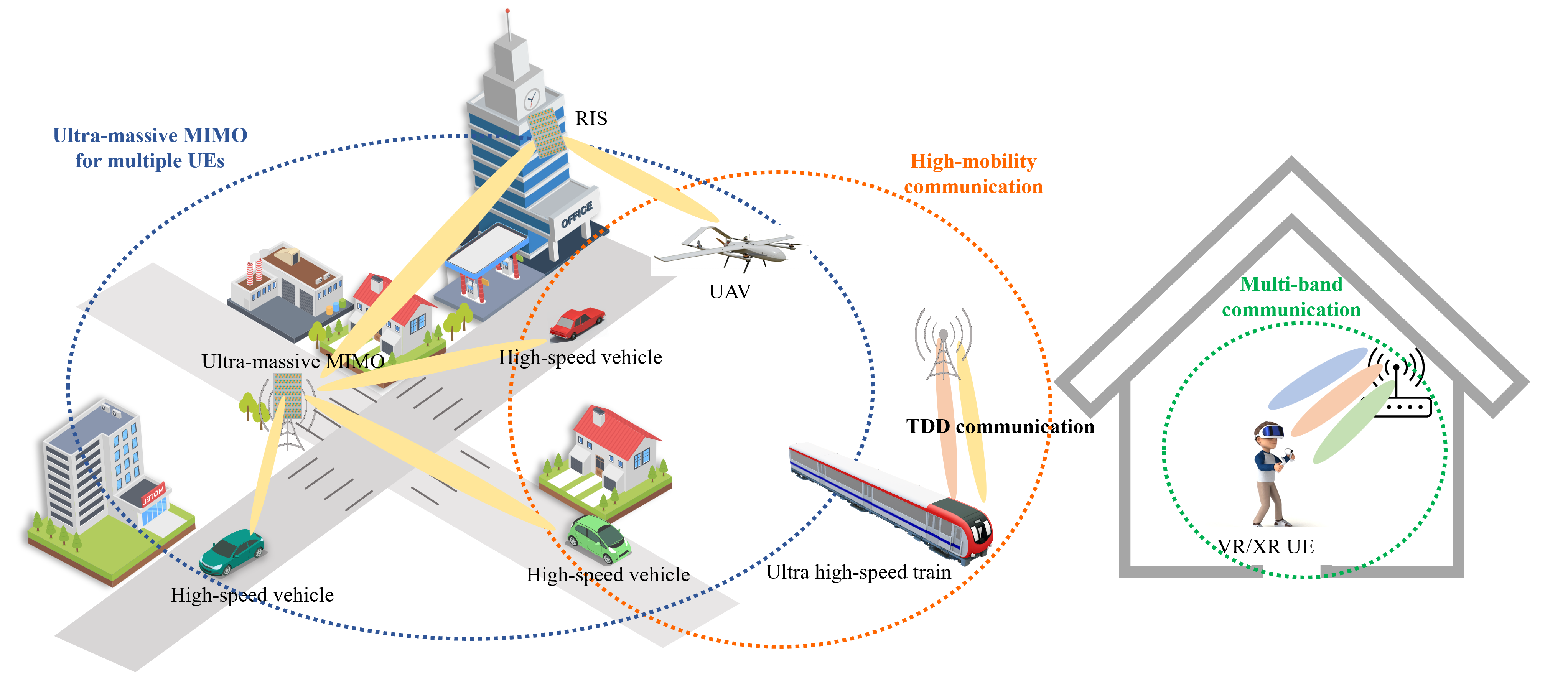} 
\caption{Emerging scenarios of 6G that calls for multi-domain channel extrapolation.} 
\label{scenario}\end{figure*}
\begin{table*}\label{Convention_algorithm}
\caption{Summary of channel extrapolation methods and their limitations for multi-domain channel extrapolation}
\centering
\begin{tabular}{ccc} 
\hline
Name of
  algorithm          & Principles for channel
  extrapolation                                                                                               & \begin{tabular}[c]{@{}c@{}}Limitations for multi-domain channel extrapolation\end{tabular}  \\ 
\hline
Auto-regressive (AR)                           & \begin{tabular}[c]{@{}c@{}}Extrapolated CSI assumed to be a linear combination \\of the known CSI, plus a stochastic term or noise\end{tabular}                                         & Poor performance in complex environments                                                                     \\ 
\hline

\begin{tabular}[c]{@{}c@{}}Parametric  \\channel-based\end{tabular} 
 & \begin{tabular}[c]{@{}c@{}}Quasi-static assumption of key channel parameters\end{tabular}                                              & \begin{tabular}[c]{@{}c@{}}Difficulty in parameter estimation, invalid for\\ dynamic scenarios\end{tabular}           \\ 
\hline
RNN-series models                         & \begin{tabular}[c]{@{}c@{}}Capturing relationship between CSIs in each domain,\\updating at each step\end{tabular}                                   & \begin{tabular}[c]{@{}c@{}}  Error accumulation\end{tabular}             \\ 
\hline
CNN                          & \begin{tabular}[c]{@{}c@{}}Extracting spatial features of CSIs\end{tabular}                                             &  Inability
of capturing long-range dependency                                                          \\ 
\hline

\end{tabular}
\end{table*}

The integration of artificial intelligence (AI) has significantly improved the accuracy of channel extrapolation \cite{zhang2023ai,zhang2021deep}. Deep learning models, including convolutional neural networks (CNNs) \cite{stenhammar2024comparison}, gated recurrent units (GRUs) \cite{li2024gan}, and Transformers \cite{jiang2022accurate}, have been explored in this context. However, current research faces limitations that impede practical application in mobile communication systems. First, many studies concentrate on a single domain, i.e., time-, frequency-, or space-domain channel extrapolation. The growing complexity of sixth-generation (6G) scenarios necessitates channel extrapolation in multi-domain due to two reasons. On one hand, the multi-domain channel extrapolation is more efficient and can reduce pilot overhead by leveraging the shared environmental information across domains\cite{han2021multi}. For example, the multi-domain channel extrapolation allows the system to infer downlink CSI from uplink measurements, significantly reducing overhead\cite{zhang2021deep}. On the other, the multi-domain channel extrapolation can not be considered as a single summation of channel extrapolation in multiple single domains. For example in high-mobility scenarios with multipath propagation, doubly-selective channel is common, where the time-domain and frequency-domain channel correlations are coupled\cite{wang2023road}. Considering all domains simultaneously is expected to achieve higher accuracy in CSI extrapolation. Second, existing research primarily aims to enhance extrapolation accuracy, often neglecting other important performance indicators. With the integration of AI in mobile communications, the Third Generation Partnership Project (3GPP) has begun standardization efforts to assess the benefits and costs associated with AI, including training overhead, inference latency, and computational complexity \cite{38_843}. Third, many studies inadequately address generalization performance, a crucial metric that evaluates a model's ability to perform well in new and unseen situations.

To tackle these challenges, we propose using generative artificial intelligence (GAI) for multi-domain channel extrapolation and introduce a Transformer encoder-like model for joint time-frequency-space domain extrapolation. In this paper, we first review conventional algorithms for channel extrapolation and examine their limitations in the context of multi-domain channel extrapolation. We then outline the performance requirements for effective multi-domain channel extrapolation, including extrapolation accuracy, generalization ability, and computational efficiency. Based on these criteria, we analyze the strengths and weaknesses of standard GAI models regarding multi-domain channel extrapolation. For our case study, we propose a Transformer encoder-like model customized for multi-domain channel extrapolation by removing the positional encoding module and substituting the original multi-head attention with a multilayer perceptron (MLP). Simulation results demonstrate that our model outperforms existing baseline models in terms of both extrapolation accuracy and inference speed. Through ablation studies, we found that our model's design effectively enhances both channel extrapolation accuracy and inference efficiency. Finally, we pose several open questions for the development of practical GAI-based multi-domain channel extrapolation models, addressing issues like explainability, generalization, and dataset collection.

\section{Opportunities and Challenges of GAI for Multi-Domain Channel Extrapolation}
\subsection{Limitations of Existing Channel Extrapolation Research}
Channel extrapolation serves as an efficient method for obtaining CSI with minimal overhead. Current research mainly focuses on single-domain channel extrapolation, namely time-, frequency-, and space-domain channel extrapolation. Existing techniques can be categorized into auto-regressive (AR)-based models \cite{wu2021channel}, parametric channel-based methods\cite{han2021multi}, recurrent neural network (RNN)-series models, and convolutional neural network (CNN)-based models \cite{stenhammar2024comparison}, as summarized in Table \ref{Convention_algorithm}.

AR-based models are widely used for channel extrapolation attributed to its simplicity and ease of implementation, which assumes that the unknown CSI is a linear combination of known CSI with a stochastic noise. This method has proven effective for channel extrapolation in a single domain, such as time- and space-domain. However, it struggles when applied to multiple domain channel extrapolation due to its dependence on simple linear relationships. In scenarios, such as those with extensive signal reflections or fast-moving devices, the assumption of channel linearity does not hold, where the performance of AR-based models degrade significantly\cite{wu2021channel}.

Parametric channel-based methods function on the assumption that channel characteristics can be described by a specific set of parameters. These techniques facilitate channel extrapolation across time, frequency, and spatial domains by estimating variations in these parameters. For example, in the time domain, parametric methods often use the quasi-static assumption, treating key parameters like complex amplitude, delay, and Doppler shift as constant or slowly changing. However, in multi-domain extrapolation, these methods face limitations, particularly in high-speed communication scenarios. Accurately estimating key parameters becomes challenging, reducing their effectiveness. Moreover, the quasi-static assumption does not hold in high-speed or multi-band communication systems, where channel parameters can vary significantly across different domains, further complicating channel extrapolation\cite{han2021multi}. 

RNN-series models have been utilized in both time and frequency-domain channel extrapolation\cite{stenhammar2024comparison}. In time-domain extrapolation, RNNs are capable to store and update hidden states at each time step, which enables them to capture and utilize temporal dependencies effectively. Nonetheless, a critical limitation is the accumulation of extrapolation errors. In a typical time-domain extrapolation process, the RNN sequentially predicts the CSI for future frames. Initially, the CSI for the first future frame is predicted using only the historical CSI; subsequent predictions rely on both this initial prediction and historical CSI. Consequently, any errors in the first prediction are propagated and combined with errors in further predictions, leading to rapid error accumulation. This issue becomes more pronounced when applied to multi-domain channel extrapolation, where accuracy is paramount across different domains.

CNN models have been investigated for extrapolating information across time, frequency, and spatial domains. Thanks to their powerful feature extraction capabilities, CNNs can learn the relationships between CSI across local time steps, frequency bands, and multiple antennas. However, the limited receptive field of CNNs restricts each convolutional filter to capturing only local features, which prevents the model from understanding broader, global information\cite{stenhammar2024comparison}. This inability to effectively capture long-range dependencies limits the usefulness of CNNs in multi-domain channel extrapolation tasks.

\begin{figure*}[htbp]\centering    
\includegraphics[width=1.8\columnwidth]{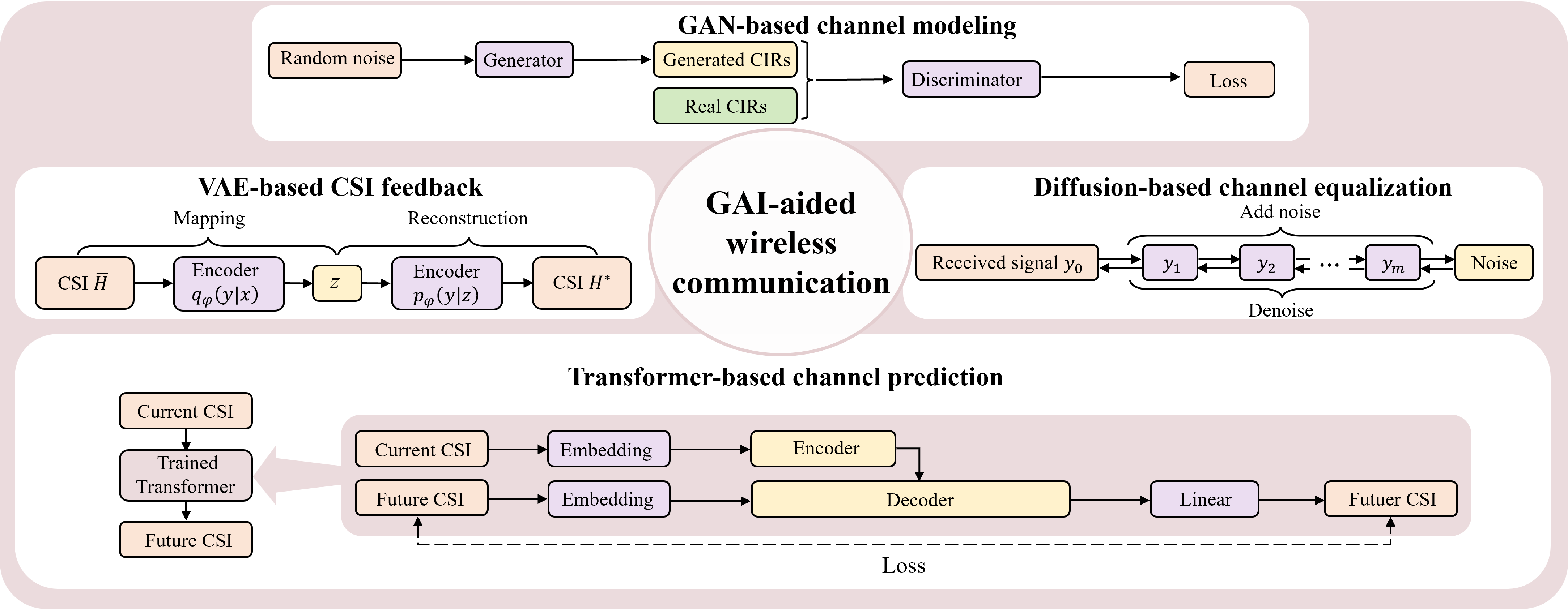} 
\caption{Illustration of typical GAI models and applications in wireless communications.} 
\label{Hol_GAI}\end{figure*}

\subsection{Key Requirements of Multi-Domain Channel Extrapolation}
6G is expected to support a wide range of emerging applications, including extended reality (XR), smart cities, and smart transportation, all of which require multi-domain channel extrapolation. For instance, as shown in Fig. \ref{scenario}, supporting communication for high-speed vehicles using ultra-massive MIMO systems necessitates joint time-space-domain channel extrapolation. Additionally, to ensure high throughput for XR applications, joint frequency-space-domain channel extrapolation is essential. This approach facilitates the acquisition of necessary CSI with manageable overhead, thereby enhancing overall system performance. To realize the full potential of 6G, the following requirements for multi-domain channel extrapolation must be fulfilled.

\subsubsection{Extrapolation Accuracy}
Extrapolation accuracy is crucial in multi-domain channel extrapolation, as it measures the difference between the extrapolated CSI and the actual CSI. Intuitively, the more precise the extrapolated CSI, the better the performance of mobile communications will be when utilizing that data.

\subsubsection{Generalization}
When evaluating multi-domain channel extrapolation performance, generalization is key to understanding how well the model can manage unseen data. Key aspects of generalization assessments include robustness and cross-domain performance. Robustness measures the model's resilience to noise and interference, ensuring stable performance even with fluctuations in noise levels. Cross-domain performance evaluates the model's ability to maintain consistent performance across various domains, such as different frequency bands, antenna configurations, and diverse channel conditions, including varying environments and user locations.

\begin{table*}[htbp]
    \caption{Strengths and weaknesses of typical GAI models for multi-domain channel extrapolation, where \CheckmarkBold and \XSolidBrush denote strength and weakness, respectively. VAE: variational auto-encoder; GAN: generative adversarial network; NF: normalizing flows.}
    \centering
    \begin{tabular}{|c|c|c|c|c|c|}
        \hline
        GAI Model &Long-range Dependency&Hidden Feature Capture&Real-time Processing&Training Instability&Computation Intensive  \\ \hline
         VAE& &\CheckmarkBold&\CheckmarkBold&\XSolidBrush&
         \XSolidBrush \\ \hline
        Diffusion Models&\XSolidBrush &\CheckmarkBold&& &\XSolidBrush \\ \hline
        GAN& &\CheckmarkBold&&\XSolidBrush&\XSolidBrush \\ \hline
        %NF& &\CheckmarkBold&&&\XSolidBrush \\ \hline
        Transformer&\CheckmarkBold &\CheckmarkBold&&&\XSolidBrush \\ \hline
    \end{tabular}
    \label{tab:Configurations}
\end{table*}

\subsubsection{Computational Complexity}
Computational complexity is a critical metric in multi-domain channel extrapolation. Important factors include algorithmic complexity, real-time processing, and scalability. Algorithmic complexity assesses both time and space complexities, where lower complexity indicates faster computation and reduced memory usage. Real-time processing examines the system's ability to handle data in real-time applications, ensuring that channel extrapolation is completed within the required timeframe. Furthermore, the model should efficiently scale to accommodate larger networks or more users without significantly increasing demands on computational resources.

%Optimizing computational complexity is vital to ensure the channel extrapolation algorithm functions effectively in real communication systems.

\subsection{Potentials and Challenges of GAI}

To achieve optimal performance in multi-domain channel extrapolation within complex scenarios, it is essential for the models to extract the deep and hidden relationships within CSI across various domains. GAI models, a branch of artificial intelligence focused on creating new and original content by learning in-depth patterns from existing datasets\cite{van2024generative,vu2024applications}, show great promise for enhancing multi-domain channel extrapolation. Research on typical GAI, such as Transformers, diffusion models, generative adversarial networks (GANs), and variational autoencoders (VAEs), in wireless communications has garnered significant interest due to its potential to improve channel estimation, support autonomous network management, generate synthetic data, and optimize overall network performance, as depicted in Fig. \ref{Hol_GAI}\cite{van2024generative, vu2024applications,Transformer}. The strengths and weaknesses of these models for multi-channel extrapolation are summarized in Table \ref{tab:Configurations}. 

\subsubsection{Transformer}
Transformers utilize self-attention mechanisms to analyze the features of input sequences and produce corresponding output sequences \cite{Transformer}. Their capacity to capture long-range dependencies makes them well-suited for physical layer tasks, including channel prediction and joint source-channel coding \cite{van2024generative}. Such capacity also makes Transformer promising for multi-domain channel extrapolation in complex scenarios. For example, in ultra-massive MIMO systems, the characteristics of channels can change significantly over time or across different spatial dimensions, thus making the channel extrapolation in joint time-antenna-domain challenging. Transformer can be trained to effectively capture the hidden features of CSI in time-antenna-domain, thereby enhancing the channel extrapolation performance. However, the training and inference processes for Transformers are computationally intensive, which may limit their applicability in resource-constrained environments, such as mobile devices.

%Transformers leverage self-attention mechanisms to understand the characteristics of input sequences and generate corresponding output sequences \cite{Transformer}. Their ability to capture long-range dependencies makes them suitable for tasks at the physical layer, such as channel prediction and joint source-channel coding\cite{van2024generative}. Transformers are particularly adept at learning long-term dependencies, a capability that is critical for channel extrapolation tasks. In ultra-massive MIMO systems, channel characteristics may undergo substantial changes over time or space. Transformers effectively model these dynamic variations and predict future channel states, making them indispensable for scenarios where capturing temporal and spatial evolution is essential. This ability significantly enhances the accuracy of extrapolation by enabling the model to identify and utilize intricate patterns that span different domains. However, training and inference of Transformer is computation-intense, which may restrict its application in multi-domain channel extrapolation in resource-constrained real-world systems, such as mobile devices.  

\begin{figure*}[htbp]\centering    
\includegraphics[width=1.6\columnwidth]{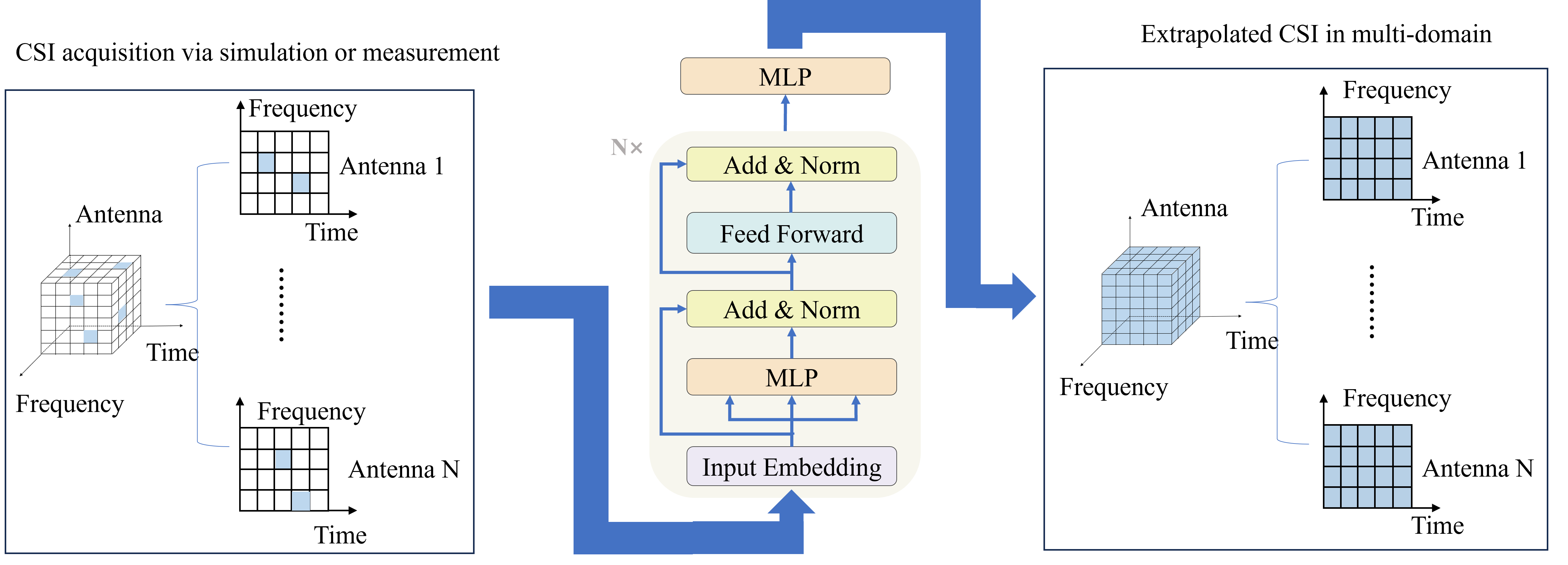} 
\caption{Pipeline of the proposed Transformer encoder-like model for multi-domain channel extrapolation.} 
\label{framework}\end{figure*}

\subsubsection{Diffusion Models}
Diffusion models generate data by progressively adding noise to samples and then learning to reverse this process to produce high-fidelity outputs. They can be trained on incomplete data in a stable manner, making them particularly effective for equalizing and modeling channels with limited training data and in noisy conditions \cite{vu2024applications}. However, the high complexity of diffusion models may limit their deployment in resource-constrained real-world systems, such as mobile devices. In addition, diffusion models face challenges in learning long-term dependencies in CSI, potentially failing to fully capture dynamic cross-domain characteristics. 

\subsubsection{Generative Adversarial Networks (GANs)}
GANs consist of a generator neural network that creates fake data and a discriminator neural network that distinguishes between fake and real data. This adversarial process drives the generator to produce more realistic data. GANs can be applied in various tasks, including channel estimation, channel modeling, CSI feedback, beamforming, and joint source-channel coding \cite{van2024generative}. However, GANs often suffer from mode collapse, where the generated channel data lacks diversity, thereby compromising the comprehensiveness of the extrapolation. 

\subsubsection{Variational Autoencoders (VAEs)}
VAEs learn the probabilistic distribution of latent space by employing an encoder-decoder architecture that reconstructs inputs and generates new data. Due to their strong reconstruction capabilities, VAEs are widely used for channel equalization, channel modeling, CSI feedback, beamforming, and joint source-channel coding \cite{vu2024applications}. The encoder and decoder of VAEs exhibit low-latency characteristics, rendering them highly promising for real-time channel extrapolation. In wireless communications, where real-time performance is a key requirement, this low-latency feature gives VAEs a distinct advantage, particularly in dynamic environments where rapid adaptation to changing conditions is necessary. However, VAEs are sensitive to noise and prone to overfitting, which can result in less robust extrapolation outcomes.

\section{GAI for Multi-domain Channel Extrapolation }
Given the ability of the Transformer to capture long-range dependencies and hidden patterns, we conducted a case study to investigate its potential for multi-domain channel extrapolation. Specifically, we propose a Transformer encoder-like framework for joint time-frequency-space-domain channel extrapolation, as depicted in Fig. \ref{framework}. This framework differentiates itself from the conventional Transformer model in two key areas.
\subsection{Encoder \& Decoder}
The encoder and decoder are critical components of the Transformer architecture. The encoder consists of input embedding, a multi-head self-attention mechanism, a feed-forward network, residual connections, and layer normalization. It transforms input sequences into fixed-dimensional vectors while capturing global information through self-attention. The decoder, while similar, includes a masked self-attention mechanism to facilitate sequence generation. Although encoder-decoder architectures are effective for complex sequence-to-sequence tasks, they tend to be computationally intensive.

Inspired by the BERT, a Transformer encoder-like GAI model for natural language processing (NLP), we adopt an encoder-only architecture for multi-domain channel extrapolation due to its efficiency in feature extraction. By using the encoder-only model, we simplify the architecture, reducing computational demands while effectively capturing channel features. This approach is particularly well-suited for tasks that require an in-depth understanding of the inputs, thus maintaining robust performance across various domains.

\subsection{Attention Mechanism \& Positional Encoding}
%Multi-head attention facilitates the capture of long-range dependencies by allowing the model to focus on different parts of the input sequence simultaneously, while positional encoding preserves the sequential order of the data, enabling contextual understanding. Attention mechanism and positional encoding together ensure that the Transformer remains effective in handling complex sequential tasks, allowing the model to pay attention to different parts of the input concurrently, meaning that important contextual relationships can be maintained even if individual token orders change. This is denoted as the permutation invariance, and is the key feature of Transformer. 

The key feature of Transformers is the multi-head self-attention mechanism, which effectively extracts semantic correlations among elements in long sequences, such as words in text or 2D patches in images\cite{Transformer}. However, self-attention is inherently permutation-invariant. Although various types of positional encoding can preserve ordering information, applying self-attention can still lead to a loss of temporal information. This loss may not be critical for semantic-rich applications like natural language processing, where reordering words often does not significantly change the meaning of the sentence\cite{DLinear}.

In the context of channel extrapolation, the permutation-invariance of the self-attention mechanism is not desirable. For instance, in time-domain channel extrapolation, the temporal CSI series often lacks intrinsic semantics. The primary focus is on modeling the temporal variations in a continuous set of CSI values. Consequently, the sequence order in temporal CSI is vital for accurately understanding the temporal corrections of the CSI sequence. Adding positional encodings to the CSI sequence may lead to the loss of original sequence information or positional context, ultimately degrading extrapolation performance.

To address these problems, we draw inspiration from recent research on time-series prediction using Transformers \cite{DLinear}. We propose to eliminate positional encoding and replace the multi-head self-attention mechanism with an MLP. This adjustment allows the model to inherently capture positional information without needing additional positional encoding \cite{tolstikhin2021mlp}. Furthermore, MLPs are much less computation-intensive than multi-head self-attention modules, resulting in reduced complexity for the proposed model.

\begin{table*}\centering\caption{Performance comparison of various models in terms of channel extrapolation accuracy and inference speed. T-D, TF-D, TD-D and TFS-D refer to time-domain, time-frequency-domain, time-space-domain and time-frequency-space-domain channel extrapolation, respectively. Proposed model+* replaces the MLP with multi-head self-attention module and adopts the positional encoding. Proposed model* denotes the proposed model with positional encoding. Proposed model+ replaces the MLP with multi-head self-attention module. In-distribution evaluation indicates that the model are trained and tested using the same dataset, i.e., dataset A. Out-of-distribution evaluation indicates that the models are trained and tested using dataset A and B, respectively.}
% Please add the following required packages to your document preamble:
% \usepackage{multirow}
\begin{tabular}{cc|ccccccccc}
\hline
\multicolumn{2}{c|}{\multirow{2}{*}{}} & \multicolumn{1}{c|}{\multirow{2}{*}{\begin{tabular}[c]{@{}c@{}}Inference speed \\ (Samples/second)\end{tabular}}} & \multicolumn{4}{c|}{SGCS} & \multicolumn{4}{c}{NMSE (dB)} \\ \cline{4-11} 
\multicolumn{2}{c|}{} & \multicolumn{1}{c|}{} & T-D & TF-D & TS-D & \multicolumn{1}{c|}{TFS-D} & T-D & TF-D & TS-D & TFS-D \\ \hline
\multicolumn{1}{c|}{\multirow{7}{*}{\begin{tabular}[c]{@{}c@{}}In-distribution \\ performance\end{tabular}}} & UNet & 205.89 & 0.7399 & 0.7399 & 0.7311 & 0.7354 & -3.7997 & -3.7987 & -3.6190 & -3.7078 \\
\multicolumn{1}{c|}{} & GRU {[}6{]} & 258.62 & 0.8388 & 0.8387 & 0.8268 & 0.8326 & -6.2953 & -6.2923 & -5.8833 & -6.0724 \\
\multicolumn{1}{c|}{} & Transformer {[}13{]} & 159.87 & 0.8280 & 0.8280 & 0.8152 & 0.8213 & -5.5874 & -5.5872 & -5.2468 & 5.4085 \\
\multicolumn{1}{c|}{} & Proposed Model & \textbf{388.21} & \textbf{0.8754} & \textbf{0.8755} & \textbf{0.8633} & \textbf{0.8691} & \textbf{-7.1007} & \textbf{-7.1006} & \textbf{-6.5512} & \textbf{-6.8040} \\
\multicolumn{1}{c|}{} & Proposed Model+* & 282.89 & 0.8255 & 0.8255 & 0.8135 & 0.8193 & -5.6861 & -5.6865 & -5.3174 & -5.4918 \\
\multicolumn{1}{c|}{} & Proposed Model* & 386.52 & 0.8481 & 0.8480 & 0.8388 & 0.8432 & -6.7523 & -6.7522 & -6.2007 & -6.4546 \\
\multicolumn{1}{c|}{} & Proposed Model+ & 283.10 & 0.8015 & 0.8015 & 0.7883 & 0.7947 & -5.0267 & -5.0277 & -4.8591 & -4.9526 \\ \hline
\multicolumn{1}{c|}{\multirow{7}{*}{\begin{tabular}[c]{@{}c@{}}Out-of-distribution\\ performance\end{tabular}}} & UNet & 206.13 & 0.6982 & 0.6983 & 0.6881 & 0.6933 & -3.0071 & -3.0053 & -2.8650 & -2.9387 \\
\multicolumn{1}{c|}{} & GRU {[}6{]} & 259.02 & 0.7482 & 0.7483 & 0.7381 & 0.7433 & -3.7071 & -3.7053 & -3.5650 & -3.6387 \\
\multicolumn{1}{c|}{} & Transformer {[}13{]} & 160.70 & 0.7628 & 0.7628 & 0.7518 & 0.7572 & -4.0362 & -4.0365 & -3.8071 & -3.9167 \\
\multicolumn{1}{c|}{} & Proposed Model & \textbf{384.45} & \textbf{0.8247} & \textbf{0.8248} & \textbf{0.8155} & \textbf{0.8000} & \textbf{-5.4605} & \textbf{-5.4607} & \textbf{-5.2629} & \textbf{-5.3560} \\
\multicolumn{1}{c|}{} & Proposed Model+* & 281.09 & 0.7789 & 0.7788 & 0.7696 & 0.7747 & -4.2970 & -4.2977 & -4.0487 & -4.1642 \\
\multicolumn{1}{c|}{} & Proposed Model* & 381.23 & 0.8083 & 0.8083 & 0.8010 & 0.8047 & -5.0326 & -5.0335 & -4.8730 & -4.9428 \\
\multicolumn{1}{c|}{} & Proposed Model+ & 284.36 & 0.7552 & 0.7553 & 0.7528 & 0.7557 & -3.9462 & -3.9496 & -3.6651 & 3.8058 \\ \hline
\end{tabular}
\label{speed_SGCS_NMSE}
\end{table*}

\subsection{Performance Tests and Ablation Experiments}

\subsubsection{CSI Data}
To evaluate the performance of the proposed model, we adopt the cluster-delay-line (CDL) model, which adheres to 3GPP standard, to generate CSI dataset. According to the requirements in 3GPP 38.843 for generalization performance evaluation\cite{38_843}, we generate dataset A and B by choosing different simulation settings. The central frequency is 28 GHz. Tx and Rx are equipped with 4 and 2 antennas, respectively. The subcarrier spacing of 15 kHz, and the delay spread is randomly selected between [50,300] ns. The maximum Doppler frequency deviation of dataset A and B are set to be 1400 Hz and 1500 Hz, respectively. 

\subsubsection{Model Training}
Our proposed models ($N=6$) were trained on NVIDIA GeForce RTX 4090 GPUs to minimize the mean square error (MSE) between the extrapolated multi-domain CSI and the ground truth of the multi-domain CSI. The AdamW optimizer was utilized with a maximum learning rate of 0.0003, a batch size of 32, and a weight decay of 0.01. Training spanned 100 epochs, employing the 1 cycle learning rate annealing policy to dynamically adjust the learning rate. For testing, we used the NVIDIA GeForce GTX 1050 Ti, a more accessible GPU, to evaluate model performance in real-world scenarios. This choice ensures that our evaluation reflects typical deployment scenarios.

The dataset A is divided such that 80\% is assigned to the training set, while 10\% and 10\% are allocated to the validation and test sets, respectively. In each epoch of the training phase, the model generates predictions through forward propagation and calculates the loss relative to the true targets. Subsequently, the model parameters are adjusted using backpropagation to optimize the loss function. During this phase, the validation set is employed periodically to evaluate model performance, helping to prevent overfitting and optimize hyper-parameters. At the end of the training process, the optimal model weights are saved and loaded for inference, allowing predictions to be made without gradient computation. For in-distribution evaluation, the proposed model and benchmark models are trained and tested using dataset A. For out-of-distribution evaluation, the proposed model and benchmark models are trained and tested using dataset A and B, respectively. 

%We trained the proposed models and compared models on high-performance NVIDIA GeForce RTX 4090 GPUs. We implemented the AdamW optimiser\cite{loshchilov2017decoupled} using carefully tuned hyperparameters: a maximum learning rate of 0.0003, a batch size of 32, and a weight decay of 0.01. The training process for all models spanned 100 epochs. to optimise the learning process, we used the 1cycle learning rate annealing policy\cite{smith2019super} that dynamically adjusted the learning rate throughout the training process.

%During the testing phase, we turned to a more widely used GPU, the NVIDIA GeForce GTX 1050 Ti. This choice was motivated by the fact that it is more readily available for real-world applications, allowing us to evaluate our models under conditions that are more representative of real-world scenarios. This approach provides insight into the performance of each model on hardware commonly found in typical deployment environments.
\subsubsection{Results Analysis}

As outlined by 3GPP 38.843, two performance metrics are defined for AI/machine learning (ML)-enhanced CSI-related tasks: average squared generalized cosine similarity (SGCS) and normalized mean square error (NMSE) \cite{38_843}. Among these metrics, SGCS is particularly effective for assessing the relationship between vectors, making it an excellent choice for eigenvector compression. Conversely, NMSE may be more beneficial for matrix compression. We utilized both SGCS and NMSE to evaluate the accuracy of channel extrapolation comprehensively.

Table \ref{speed_SGCS_NMSE} presents a quantitative comparison of the proposed model, UNet, GRU \cite{stenhammar2024comparison}, and Transformer \cite{Transformer} in terms of inference speed and extrapolation accuracy across different domains, including time-domain (T-D), time-frequency-domain (TF-D), time-space-domain (TS-D), and time-frequency-space-domain (TFS-D). The table indicates that in in-distribution evaluation, the proposed framework achieves superior inference speed and extrapolation accuracy compared to the other models. We further carried out out-of-distribution evaluation to test the generalization performance of the proposed model. Although all the models experience performance degradation in out-of-distribution evaluation, the proposed model still outperforms other benchmarks, demonstrating its superior generalization over existing models.

In Fig. \ref{speed_SGCS_NMSE}, the ablation experiments are carried out to evaluate the effects of eliminating positional encoding and replacing multi-head self-attention with an MLP. The proposed model+* replaces the MLP with a multi-head self-attention module and employs positional encoding, while Proposed Model* denotes the proposed model with positional encoding only. Proposed Model+ replaces the MLP with a multi-head self-attention module. The proposed model outperforms proposed model* in channel extrapolation accuracy with comparable inference speed, indicating that positional encoding leads to a loss of original sequence information or positional context. Proposed model+ exhibits the poorest performance in both inference speed and channel extrapolation accuracy, suggesting that the multi-head self-attention module is computationally intensive and less effective at capturing correlations in CSI within multi-domain settings. The limitations of the multi-head self-attention module in multi-domain channel extrapolation are mitigated by the addition of positional attention in the proposed model+ with a slight improvement in terms of extrapolation accuracy. In summary, the findings from the ablation experiments echo the analysis of the attention mechanism and positional encoding in the previous section.

\section{Future Research Directions}
\subsection{Explainability of GAI}
As GAI-driven model are expected to enhance the performance of wireless communication systems, the need for explainability—the ability to understand and interpret the decisions made by GAI models—has become an essential consideration for several reasons. First, explainability fosters trust among network operators, engineers, and end users. When stakeholders can comprehend how AI algorithms arrive at their decisions, they are more likely to embrace these technologies. Trust is particularly vital in telecommunications, where system reliability and performance are crucial. Furthermore, when AI systems encounter errors or produce unexpected outcomes, explainability allows users to trace the decision-making process back to its origins, which is essential for improving model accuracy and addressing issues in real-time network management.
\subsection{Generalization Performance of GAI}
Generalization performance is a crucial indicator of how effectively GAI models can operate across a wide array of scenarios. It reflects the model's ability to apply what it has learned during training to unseen data, ensuring that it can maintain high levels of accuracy and reliability in diverse conditions. This capability is particularly important in fields such as wireless communications, where conditions can vary significantly due to factors like changes in the environment, user behavior, and network configurations. 3GPP Technical Report 38.843 highlights the significance of generalization performance, emphasizing that GAI models used in CSI related tasks must not only perform well during training but also adapt successfully to new, real-world data\cite{38_843}. This ensures that the models developed are robust and versatile, reducing the risk of overfitting to specific training datasets and allowing them to deliver consistent performance in practical applications.
\subsection{Dataset and Testbeds}
Most current research on channel extrapolation, even wireless communication, are based on simulation. The applicability of the corresponding algorithms and frameworks are doubtable in practical mobile networks. Measurement CSI is valuable resource for companies, and is difficult for the academia to acquire. Efforts have been made to generate practical channel data by using simulators, of which NYUSIM\footnote{https://wireless.engineering.nyu.edu/nyusim/} BUPTCMCCCMG-IMT2030\footnote{https://hpc.bupt.edu.cn/dataset-public/datasets/29} and SEU-PML-6GPCS\footnote{https://www.wjx.top/vm/OtYwTui.aspx} are the most typical cases. Several open source CSI data are public accessible, but not comprehensive. Specifically, CSI data for ultra-massive MIMO, high speed and multi-band systems are extremely difficult to measure and are absent for research. How to acquire high quality CSI data with low cost remains an open challenge. GAIs, especially GAN, are promising candidate to generate synthetic CSI data that closely resemble measurement CSI data. As mentioned in the previous section, the key challenge lays in training GAN-based model.

%\subsection{Reconfigurable intelligent surface (RIS)-assisted communication}
%RIS can adaptively manipulate the phase and amplitude of the incident electromagnetic signal and reflect it to the desired direction, which enhances the quality of the signal at the receiver and hardens the wireless channel to achieve spectrum-efficient and energy-efficient networks \cite{shlezinger2021dynamic}. Nonetheless, in order to fully exploit the potential performance gain of RIS, it is necessary to obtain the explicit CSI over the RIS-assisted networks. Due to the structural characteristics of RIS, the size of its cascaded channels is usually proportional to the number of RIS elements, which leads to a prohibitive pilot consumption. Moreover, in complex physical environments and high-frequency band communication scenarios, conventional signal processing methods perform poorly in dealing with non-ideal channel conditions. In fact, the high dense placement of RIS elements implies strong correlation between the channels on adjacent elements. Therefore, channel extrapolation can exploit the mapping relationship between compressed and complete channels to infer complete channel information from partial channel observations. This improves the accuracy of RIS channel estimation, thus guaranteeing the effective deployment and performance enhancement of RIS in real communication networks.
\section{Conclusions}

6G is anticipated to provide services in complex scenarios, making multi-domain channel extrapolation essential for acquiring accurate CSI with minimal overhead. In this paper, we introduce the concept of multi-domain channel extrapolation along with its critical performance requirements. We explore both the potential benefits and challenges of integrating GAI-based models for effective multi-domain channel extrapolation. In the case study, we propose a novel Transformer encoder-like model tailored for multi-domain channel extrapolation. Simulation results demonstrate that our proposed model outperforms existing baseline models in terms of extrapolation accuracy and inference speed. Through ablation studies, we discover that removing the positional encoding module and replacing the original multi-head attention mechanism with an MLP effectively enhance both channel extrapolation accuracy and inference speed. Finally, we emphasize the need to address issues related to explainability, generalization, and dataset collection to ensure the practical application of GAI in multi-domain channel extrapolation.

\bibliographystyle{IEEEtran}
\bibliography{Joint_extrapolation}

\end{document}